\documentclass{nature}
\usepackage{graphicx}%
\usepackage{multirow}%
\usepackage{amsmath,amssymb,amsfonts}%
\usepackage{amsthm}%
\usepackage{mathrsfs}%
\usepackage[title]{appendix}%
\usepackage{xcolor}%
\usepackage{textcomp}%
\usepackage{manyfoot}%
\usepackage{booktabs}%
\usepackage{algorithm}%
\usepackage{algorithmicx}%
\usepackage{algpseudocode}%
\usepackage{listings}%
\usepackage{txfonts}
\usepackage{hyperref}
\usepackage{tablefootnote}
\usepackage[T1]{fontenc}
\usepackage{pdfpages}

\title{Insights into star formation and dispersal from the synchronisation of stellar clocks} 

\author{Núria Miret-Roig$^{1}$, João Alves$^{1}$, David Barrado$^{2}$, Andreas Burkert$^{3,4,5}$, Sebastian Ratzenböck$^{1,6}$  \& Ralf Konietzka$^{4,7,8}$ }

\begin{document}

\maketitle

\begin{affiliations}
\item University of Vienna, Department of Astrophysics, Türkenschanzstraße 17, 1180 Wien, Austria; e-mail: \url{nuria.miret.roig@univie.ac.at}
\item Centro de Astrobiología (CSIC-INTA), Depto. de Astrofísica, ESAC Campus, Camino Bajo del Castillo s/n, 28692, Villanueva de la Cañada, Madrid, Spain
\item Universitäts-Sternwarte, Ludwig-Maximilians-Universität München, Scheinerstr. 1, 81679 Munich, Germany
\item Max-Planck Institute for Extraterrestrial Physics, Giessenbacherstr. 1, 85748 Garching, Germany
\item Excellence Cluster ORIGINS, Boltzmannstrasse 2, 85748 Garching, Germany
\item University of Vienna, Research Network Data Science at Uni Vienna, Kolingasse 14-16, 1090 Wien, Austria
\item Ludwig-Maximilians-Universität München, Geschwister-Scholl Platz 1, 80539 Munich, Germany
\item Harvard University Department of Astronomy, Center for Astrophysics | Harvard \& Smithsonian, 60 Garden St. Cambridge, MA 02138, USA
\end{affiliations}

\begin{abstract}

Age is one of the most fundamental parameters of stars, yet it is one of the hardest to determine as it requires modelling various aspects of stellar formation and evolution. When we compare the ages derived from isochronal and dynamical traceback methods for six young stellar associations, we find a systematic discrepancy. Specifically, dynamical traceback ages are consistently younger by an average of $\langle\Delta_{\rm Age}\rangle = 5.5 \pm 1.1$ Myr. We rule out measurement errors as the cause of the age mismatch and propose that $\Delta_{\rm Age}$ indicates the time a young star remains bound to its parental cloud before moving away from its siblings. In this framework, the dynamical traceback “clock” starts when a stellar cluster or association begins to expand after expelling most of the gas, while the isochronal “clock” starts earlier when most stars form. The difference between these two age-dating techniques is a powerful tool to constraint evolutionary models, as isochronal ages cannot be younger than dynamical traceback ages. Measuring the $\Delta_{\rm Age}$ accurately and understanding its variations across different environments will provide further information on the impact of local conditions and stellar feedback on the formation and dispersal of stellar clusters. 
\end{abstract}


Ages are crucial to understanding most astrophysical processes. In particular, the ages of young stars are fundamental for studying the formation and early evolution of stars and planets \cite{Drazkowska+2023, Manara+2023}, establishing timescales for protoplanetary disk dissipation \cite{Williams+2011}, determining stellar masses from O-stars to substellar objects (brown dwarf, bound planets and free-floating planets \cite{Miret-Roig+2023}), and studying the recent star formation history of the solar neighbourhood \cite{Zucker+2023}. However, estimating precise stellar ages for individual stars (and even clusters) is complicated since different methodologies work best for different ages and masses, and the results are often strongly model-dependent \cite{Mermilliod2000, Soderblom2010-lp, Soderblom+2014, Barrado2016}. Stellar clusters and associations containing co-eval stars constitute excellent benchmarks for age-dating methods because they represent snapshots in evolution for stellar ensembles across a large range of masses (at a given metallicity). Absolute ages for young stellar associations can be obtained from evolutionary models (e.g., isochrone fitting, lithium depletion boundary, and asteroseismology) or the analysis of their kinematics (e.g., dynamical tracebacks or assuming linear expansion). 

Isochrone fitting is one of the most common techniques to determine stellar ages. The excellent photometry and distances from \textit{Gaia} \cite{GaiaColVallenari+2022} have provided systematic isochronal ages for hundreds of previously known clusters and associations and thousands of Galactic stellar ensembles that have been recently identified \cite{Bossini+2019, Cantat-Gaudin+2020, Dias+2021, Li+2022}. However, isochrones are based on stellar evolutionary models, strongly dependent on the complex physical processes included, and are particularly uncertain for young, pre-main sequence stars and low-mass stars \cite{Baraffe+2002, Luhman+2012, Allard+2014}, where different families of models still have discrepancies of about a factor of two. In addition, young stars are photometrically variable (due to activity) and are affected by interstellar extinction, all of which make isochronal age dating problematic \cite{Jeffries+2014, Jeffries+2021, Binks+2022}. An alternative, potentially model-independent technique known as the lithium depletion boundary (LDB, for which there is a sharp gap in the luminosity between fully convective Li-rich and Li-poor stars) can yield precise ages and represent excellent benchmarks to test evolutionary models \cite{Basri+1996, Stauffer+1998, Barrado+1999b, Manzi+08, Binks+14, Binks+2021, Galindo-Guil+2022}. However, detecting LDBs requires many hours of spectroscopic observing time on large telescopes and this technique is only effective for clusters with ages $>20$Myr.

Dynamical traceback ages constitute another model-independent method that can be used to complement isochronal ages. This technique assumes that an expanding group of unbound stars formed when they occupied a minimum volume of space \cite{Blaauw1964, Miret-Roig+2020b}. These ages denote the traceback time required for the system to reach its minimum volume configuration. Precise 3D velocities for individual stars are critical to tracing the trajectories of stars back in time, and this has been a limiting factor for many years \cite{Brown+1997a, Ortega+2002, delaReza+2006, Ducourant+2014, Miret-Roig+2018}. Thanks to \textit{Gaia} astrometry and complementary spectroscopic surveys such as APOGEE \cite{Majewski+2017}, we can now measure 3D velocities with precisions below 1~km/s. This allows us to calculate stellar orbits a few tens of Myr back in time with precisions of a few parsecs. The availability of high-precision kinematic and positional measurements has recently revived interest in traceback analyses \cite{Miret-Roig+2020b, Miret-Roig+2022b, Squicciarini+2021, Kerr+2022b, Kerr+2022a, Galli+2023, Couture+2023, Quintana+2023}. 

Historically, there have been significant age discrepancies, of more than 50\%, in the ages determined in young stellar clusters and associations \cite{Soderblom2010-lp, Soderblom+2014, Barrado2016}. This was due to a combination of contaminated samples and significant uncertainties in the observables and the models. Since the beginning of the \textit{Gaia} era, we have censuses of co-eval stars with an unprecedented degree of completeness and low contamination. Additionally, \textit{Gaia} and complementary spectroscopic surveys have significantly reduced the uncertainties in the 6D phase space parameters (positions and velocities) of many stars producing dynamical traceback ages more accurately than in the past. These recent dynamical traceback ages have narrowed the gap between ages obtained from evolutionary models and dynamical tracebacks \cite{Miret-Roig+2020b, Galli+2023}, enabling a detailed comparison of these two independent methods for the first time.

\begin{figure*}
   \centering
   \includegraphics[width = 0.49\columnwidth]{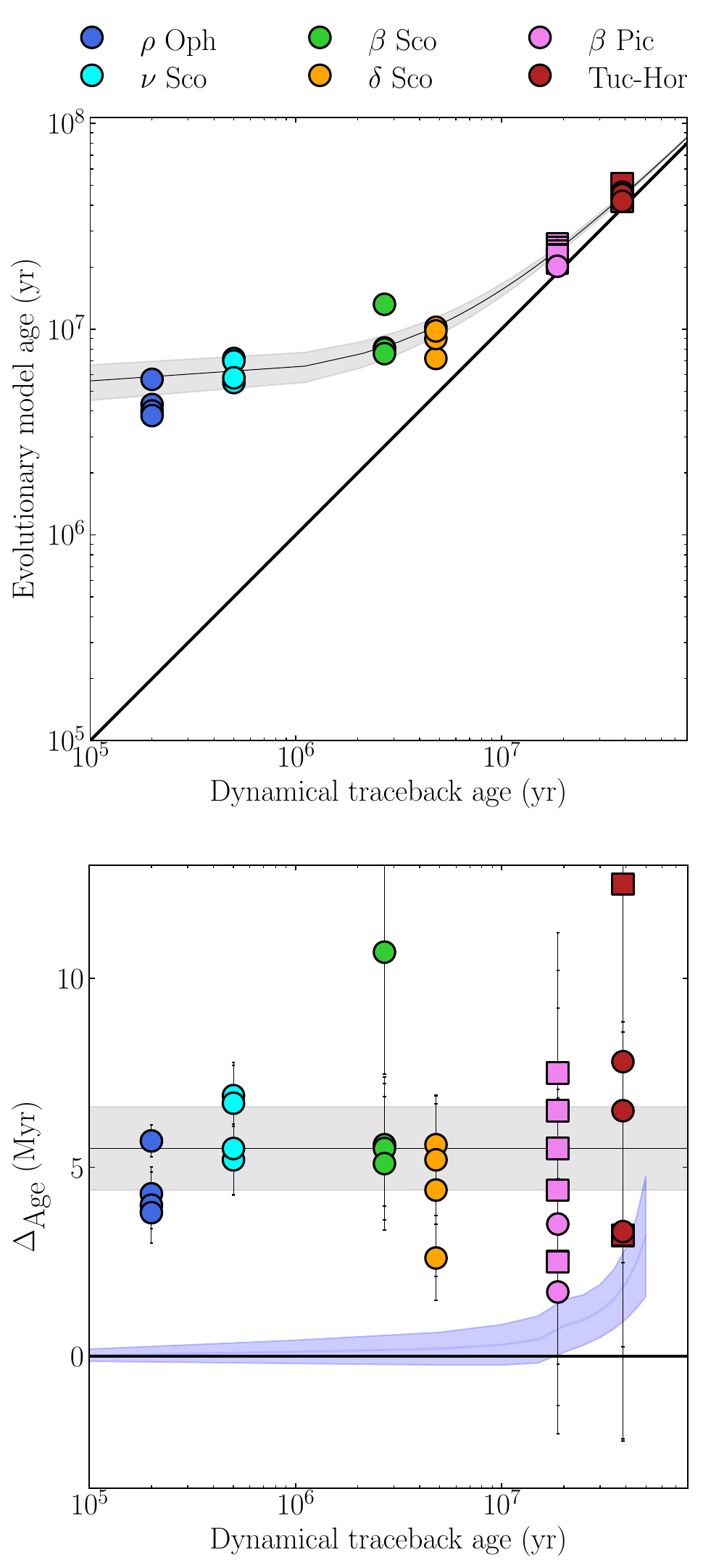}
   \caption{Systematic offset between ages from evolutionary models and dynamical traceback ages. The two panels show the one-to-one relation (top) and the absolute age difference defined as $\Delta_{\rm Age}$ = evolutionary models -- dynamical traceback age (bottom). The figure shows ages from isochrone fitting (circles) and lithium (squares). Marker colours represent different associations indicated in the legend. The position and error of the data points represent the ages and uncertainties from the studies listed in Supplementary Table~1.
   The zero offsets (thick line), the mean offset of $5.5\pm1.1$~Myr (thin line), and the standard deviation (grey shaded areas) are indicated. The blue shaded area represents the mean and standard deviation of the bias in dynamical traceback ages measured in mock simulations (see Methods). We included a $2\cdot 10^5$~yr offset to all the dynamical ages on the x-axis for visualisation purposes.}
    \label{fig:deltaT}%
    \end{figure*}

In Figure~\ref{fig:deltaT}, we compare ages from evolutionary models (isochrone fitting and lithium) and dynamical traceback ages for six benchmark associations where homogeneous and accurate dynamical traceback ages are available (see Methods section). We find a mean systematic offset, $\langle\Delta_{\rm Age}\rangle$ = $\langle$evolutionary models -- dynamical traceback age$\rangle$, of $5.5~\pm1.1$~Myr  and of $5.7~\pm1.1$~Myr when excluding $\rho$~Oph, the youngest group for which we did not find signs of expansion yet. This offset is evident despite the still large uncertainties on the age determinations with different techniques and samples. The correction of errors and biases on the current age determinations could reduce the $\Delta_{\rm Age}$ we measure, but we argue that they can not completely account for it.

Despite the improvements in observations and models, significant uncertainties still affect age determinations. Since the methods to determine stellar ages are complicated functions of the observables, it is challenging to propagate them to the final age uncertainty. Estimating and including the systematic uncertainties of evolutionary models in the final age determination is difficult, if not impossible. We included studies using different techniques (isochrone fitting and lithium) to illustrate the variability of ages within a single region, denoting the uncertainties of ages from evolutionary models. We estimated the uncertainties in the dynamical traceback ages by simulating a mock association and convolving it with realistic observational errors (see the Methods section). Uncertainties introduced by methodological choices, such as the Galactic potential and the associations' size estimator, are smaller than the uncertainties in the 3D velocities and the sample selection \cite{Miret-Roig+2020b, Couture+2023}. The errors on the dynamical traceback ages are $\lesssim1-2$~Myr for associations younger than 40~Myr. The $\Delta_\textup{Age}$ we observe in Figure~\ref{fig:deltaT} (and report in Table~\ref{tab:properties}) is larger than the scatter of ages from different evolutionary models and larger than the errors from dynamical traceback ages, indicating that uncertainties alone cannot explain this discrepancy. This is especially true for the youngest groups ($\rho$~Oph, $\nu$~Sco, $\beta$~Sco and $\delta$~Sco).

To investigate the origin of the $\Delta_\textup{Age}$, it is necessary to precisely define the \textit{birth time} of the cluster (the time the \textit{clock} starts) for evolutionary models and dynamical traceback ages. The initial time for evolutionary models is difficult to establish since the early stages of core formation and collapse are very quick ($\lesssim10^5$ yr) and hard to observe (due to large amounts of extinction). Some authors have suggested that stars are born when most of the material in the envelope has collapsed onto the disk and the central protostar becomes observable at infrared wavelengths \cite{Stahler+1983, Palla+1999}, while others place the time zero a bit earlier, at the moment when the core becomes optically thick \cite{Wuchterl+2003}. In any case, the differences between these two stages should be short, given the rapidness of these initial stages.

Dynamical traceback ages measure the time since a group of stars was most concentrated. A fundamental assumption of the current traceback methods is that stars move only affected by the gravitational pull of the Galactic potential. This is a valid assumption in many cases, given the low densities of these young associations, which are currently unbound. However, it might not have always been the case if, at birth time, there was enough gas to bind the association members together. A second important assumption of this strategy is that all stars were formed nearly simultaneously ($<1$~Myr). This assumption is supported by observations of star-forming regions \cite{Preibisch2012, Kudryavtseva+2012, Ratzenbock+2023} and the colour-magnitude diagrams of the samples considered in this study. However, small gradients (of up to 6~Myr) cannot yet be ruled out in star-forming regions spread out in large areas \cite{Krumholz+2019}. 

\begin{figure*}
   \centering
   \includegraphics[width = \textwidth]{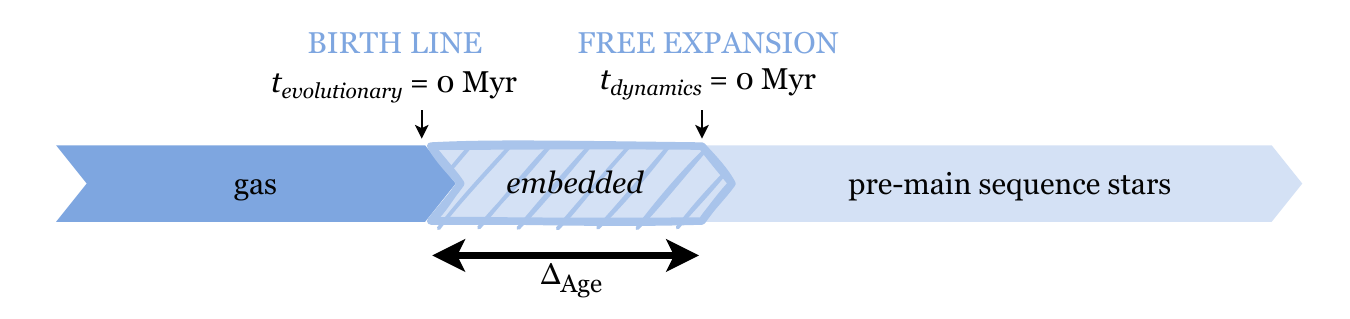}
   \caption{Diagram representing the different phases of the formation of stellar clusters and associations. Ages from evolutionary models measure the time since protostars have accreted most of their final mass and become pre-main sequence stars. Dynamical traceback ages measure the time since the association becomes unbound and starts to expand. The offset between these two techniques ($\Delta_{\rm Age}$) measures the timescale of the embedded phase, during which stars are still bound to the parent gas cloud.
   }
    \label{fig:SF_diagram}%
    \end{figure*}

Observations of unbound and expanding young clusters and associations \cite{Kuhn+2019} prove that gas removal must happen at early stages. However, this process might not be instantaneous and it has been proposed as a reason to explain the offset between isochrones and dynamical traceback ages \cite{Squicciarini+2021, Miret-Roig+2022b, Kerr+2022a}. Young stars only start moving in free trajectories in the Galactic potential after dissipating a significant fraction of the parent gas. In Figure~\ref{fig:SF_diagram}, we illustrate a timeline with our interpretation of the $\Delta_\textup{Age}$ measured in this study. Extragalactic observations have recently found similar constraints for the gas embedded phase (1--5~Myr) on larger scales (of about 100 pc) and with different strategies \cite{Kruijssen+2019, Chevance+2020, Demachi+2023}. In addition, recent star formation simulations have found that the parent gas of clusters and OB associations dissipates in 2--7 Myr \cite{Dobbs+2022, Guszejnov+2022, Jeffreson+2023}.

\begin{figure*}
   \centering
   \includegraphics[width = 0.6\columnwidth]{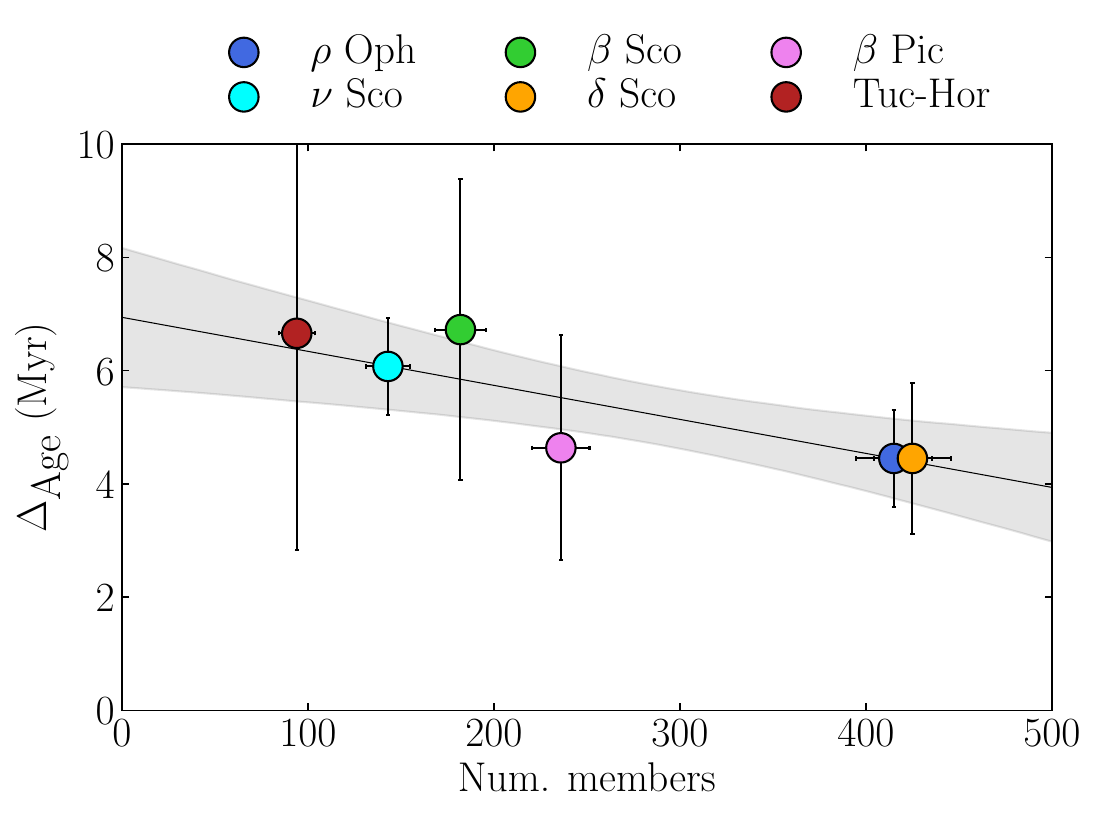}
   \caption{$\Delta_\textup{Age}$ as a function of the number of association members. The data points and errors are listed in Table~\ref{tab:properties} under the column $\Delta_\textup{Age All}$ and $N$. The black line shows the best fit and the shaded area the $1\sigma$ uncertainties.}
    \label{fig:deltaT_numMembers}%
    \end{figure*}

We have investigated the dependence of the $\Delta_\textup{Age}$ as a function of the number of members in the association (see  Figure~\ref{fig:deltaT_numMembers}), which is a relatively easy metric to obtain. The decreasing $\Delta_\textup{Age}$ with increasing association mass suggests that regions with numerous feedback events dissipate the gas more quickly. This trend is favoured by a Bayesian linear fit that considers uncertainties in the $\Delta_\textup{Age}$ (see Methods section). This relation is consistent with extragalactic observations \cite{Chevance+2022} and was predicted by hydrodynamical simulations where the gas is expelled by photoionisation radiation and stellar winds \cite{Dinnbier+2020c}. Other studies have found the timescale for gas dissipation to depend on the initial mass, compactness of the cluster, presence of OB stars and supernovae, core radius, star formation efficiency, or the velocity dispersion \cite{Goodwin+2006, Krause+2016, Dinnbier+2020a, Dinnbier+2020b}. Investigating the dependence of the $\Delta_\textup{Age}$ as a function of these parameters requires precise parameter estimates, which are currently unavailable and are left for future work.

The results of this work suggest that the $\Delta_\textup{Age}$ constitutes an observational measurement of the duration of the embedded phase and the timescale of gas dissipation. Interestingly, the mean $\langle\Delta_{\rm Age}\rangle = 5.5\pm1.1.$~Myr is similar to the crossing time of the typical initial conditions of star formation. For instance, when considering an initial dispersion in positions ($\sigma_\textup{pos}$) of 10~pc and an initial dispersion in velocities ($\sigma_\textup{vel}$) of 2~km/s, which aligns with our observations of young stellar associations studied here, the crossing time is $t_C = \sigma_\textup{pos}/\sigma_\textup{vel}\sim 5$~Myr. This coincidence might indicate that the gas removal timescale is of the order of the dynamical crossing time. In this context, we can speculate that the $\Delta_\textup{Age}$ increase for less populated associations might be due to their lower density, enlarging the dynamical crossing timescale. So, lower-mass associations might form in less dense molecular environments with larger crossing timescales.

More precise ages of stellar clusters and associations in different environments are needed to confirm the connection between the $\Delta_\textup{Age}$ and the timescale for gas dissipation. For this kind of study, identifying distinct kinematically coherent coeval groups in large star-forming regions is vital \cite{Ratzenbock+2023, Kerr+2021, Prisinzano+2022}. However, only in the solar vicinity can we have the best precision in 6D phase space coordinates, reducing the number of regions where this technique is currently applicable. Large and precise ($<1$~km/s) 3D velocity catalogues are necessary to identify robust censuses of stars (with high completeness and low contamination) and compute precise stellar orbits. These observational requirements determine which associations are better candidates for traceback ages depending on parameters such as the distance, the peculiar motion of the region, and the coverage of extensive spectroscopic surveys. We emphasise the importance of considering all the assumptions and uncertainties related to age determination, including the physics enclosed in models, the sample selection, and the need for homogeneous and precise radial velocity samples. Extensive spectroscopic surveys like APOGEE, WEAVE \cite{Dalton+2016} and 4MOST \cite{deJong+2019} are essential to complement the excellent astrometry of \textit{Gaia}. Finally, the difference between isochrone and dynamical ages naturally sets a lower limit for the ages that evolutionary models provide: they cannot be younger than dynamical ages. This can be an unexplored powerful instrument to test evolutionary models.

\begin{table*}
    \begin{center}
    \caption{Properties of the young stellar associations considered.}
    \begin{tabular}{l|c|c|c|c|c|c|c}
    \hline
    \hline
    Association       & $N$     & $d$   & DT age & Isoc. age & $\Delta_{\rm Age~All}$ & $\Delta_{\rm Age~Isoc}$ & $\Delta_{\rm Age~Lit}$ \\
                 &         & (pc)  & (Myr)  & (Myr) & (Myr)  & (Myr)     \\
    \hline
    $\rho$~Oph   & $415\pm 20$ \cite{Miret-Roig+2022b} & 139   & 0.0$\pm$0.3          \cite{Miret-Roig+2022b} & 3.8--5.7   & $4.4\pm1.0$ & $4.3\pm0.6$ & \\
    $\nu$~Sco    & $143\pm 12$ \cite{Miret-Roig+2022b} & 139   & 0.3$\pm$0.5          \cite{Miret-Roig+2022b} & 5.5--7.2   & $6.1\pm1.0$ & $5.2\pm0.9$ & \\
    $\beta$~Sco  & $182\pm 13$ \cite{Miret-Roig+2022b} & 153   & 2.4$\pm$1.7          \cite{Miret-Roig+2022b} & 7.6--13.2  & $6.7\pm2.6$ & $5.6\pm1.6$ & \\
    $\delta$~Sco & $425\pm 21$ \cite{Miret-Roig+2022b} & 142   & 4.6$\pm$0.6          \cite{Miret-Roig+2022b} & 7.2--10.2  & $4.5\pm1.3$ & $2.6\pm1.1$ & \\
    $\beta$~Pic  & $236\pm 15$ \cite{Miret-Roig+2020b} & ~~40  & $18.5_{-2.4}^{+2.0}$ \cite{Miret-Roig+2020b} & 20.2--24.0 & $4.6\pm2.0$ & $1.7\pm3.0$ & $5.3\pm1.9$\\
    Tuc-Hor      &~~$94\pm 10$ \cite{Galli+2023}       & ~~47  & $38.5_{-8.0}^{+1.6}$ \cite{Galli+2023}       & 41.8--46.3 & $6.7\pm3.8$ & $3.3\pm5.6$ & $7.8\pm6.6$\\
    \hline
    Mean         &         &       &                          & & $5.5\pm1.1$ & $3.8\pm1.5$ & $6.6\pm1.8$ \\
    Mean excl. $\rho$~Oph  &         &       &               &           & $5.7\pm1.1$ & $3.7\pm1.7$ &             \\
    \hline
    \hline
    \end{tabular}
    \label{tab:properties}
    \end{center}
    \noindent \textbf{Notes.} This table displays the name of the association (Col.~1), the number of candidate members and errors, $\sqrt{N}$, (Col.~2), the distance (Col.~3), the dynamical traceback age used for this study (Col.~4), the age range from isochrone fitting (only values with an uncertainty $<5$~Myr are considered) (Col.~5), the mean $\Delta_{\rm Age}$ and standard deviation using all precise literature determinations (Col.~6), the $\Delta_{\rm Age}$ from the PARSEC isochronal ages determined in this work with the same sample of stars used for the traceback analysis (Col.~7), and the mean $\Delta_{\rm Age}$ from Lithium ages (Col.~8). The two bottom rows contain the mean $\Delta_{\rm Age}$ and the mean $\Delta_{\rm Age}$ excluding the youngest group $\rho$~Oph.\\
\end{table*}

\begin{methods}

\noindent{\bf \large Sample selection}

While fitting isochrones to \textit{Gaia} absolute colour-magnitude diagrams is possible for many stellar groups, dynamical traceback ages require expensive spectroscopic observations and careful treatment of 3D velocities \cite{Miret-Roig+2020b}. We selected six young stellar associations for which there is a recent dynamical traceback age based on \textit{Gaia} astrometry plus precise ($\sim0.5$~km/s) radial velocities from ground-based observations, APOGEE and \textit{Gaia} \cite{Miret-Roig+2020b, Miret-Roig+2022b, Galli+2023}. These are benchmark associations, covering ages $<40$~Myr, with homogeneous radial velocities, and for which the dynamical traceback ages were determined with the same methodology. To minimise possible biases due to different samples of stars, we obtained an isochrone fitting age \cite{Ratzenbock+2023} with the same stars used for the traceback analysis and the PARSEC v1.2S models \cite{Marigo+17}. The results of these fits are shown in Supplementary Figure~1. The small scatter in these CMDs can be explained by various factors including differential extinction (especially likely in the youngest groups, $\rho$~Oph and $\nu$~Sco), photometric variability, multiplicity (almost all the stars are between the fitted isochrone and the equal-mass binary sequence, 0.75~mag brighter than the fit) or contamination from a nearby young association.

In Table~\ref{tab:properties}, we summarise the properties of the six associations considered in this study. We note that the number of members for the $\beta$~Pic and Tuc-Hor associations is more uncertain due to the difficulty of identifying members in low-density associations with strong projection effects due to their proximity ($\lesssim50$~pc). In these two cases, we decided to take the number of candidate members compiled from different studies in the literature, knowing that there is a degree of contamination from field stars and young members of other associations but also other members might remain unnoted. We took $\sqrt{N}$ as the uncertainty on the number of association members. Extensive surveys of precise radial velocities are crucial to revisit the censuses of these associations.

\textbf{Upper Scorpius and Ophiuchus.}
Upper Scorpius and Ophiuchus are among the youngest ($<10$~Myr) and closest (140~pc) star-forming regions to the Sun, part of the Scorpius-Centaurus complex. Despite its proximity, there has been a wide debate on their census and age \cite{Preibisch+2002, David+2019, Pecaut+2012, Rizzuto+2016, Feiden+2016}. Recently, several works have identified different sub-populations in this region \cite{Kerr+2021, Squicciarini+2021, Miret-Roig+2022b, Ratzenbock+2022, Ratzenbock+2023, Briceno-Morales+2023}, showing a complex star formation picture and providing an explanation for the previous age discrepancies. We selected four groups with the most robust membership determination when comparing the results between different studies. These are the $\rho$~Oph, $\nu$~Sco, $\beta$~Sco and $\delta$~Sco, which have dynamical traceback ages of $\lesssim5$~Myr \cite{Miret-Roig+2022b} and for which we measured isochrone ages of 4--8~Myr (see Table~\ref{tab:properties} and Supplementary Table~1).

\textbf{$\beta$ Pictoris moving group.}
The $\beta$ Pictoris moving group ($\beta$ Pic) is one of the closest young local associations in the solar neighbourhood, located at an average distance of only 40~pc from the Sun. It has an age of 20~Myr measured with isochrone fitting \cite{Barrado99, Malo+2014b, Mamajek14, Bell+2015, Ujjwal+20}, dynamics \cite{Crundall+19, Miret-Roig+2020b, Couture+2023}, and lithium depletion boundary \cite{Mentuch+08, Binks+14, Malo+2014b, Messina+16, Galindo-Guil+2022}. For this study, we took the members from Miret-Roig et al. 2020 \cite{Miret-Roig+2020b}, which have a dynamical traceback age of $18.5_{-2.4}^{+2.0}$~Myr and for which we measured an isochrone age of $20.2\pm2.2$~Myr.

\textbf{Tucana-Horologium association.}
The Tucana-Horologium association (Tuc-Hor) is located at around 47~pc, and has an age of 30--50~Myr from isochrone fitting \cite{Torres+2000, Bell+2015}, dynamics \cite{Kerr+2022b, Galli+2023}, and lithium depletion boundary \cite{Mentuch+08, Kraus+2014, Galindo-Guil+2022}. For this study, we took the members from Galli et al. 2023 \cite{Galli+2023}, which have a dynamical traceback age of $38.5_{-8.0}^{+1.6}$~Myr and for which we measured an isochrone age of $41.8\pm3.4$~Myr. 

\noindent{\bf \large Estimating the observational bias in dynamical traceback ages}

We simulated a mock association of different ages \cite{Miret-Roig+2018} to estimate the uncertainties on dynamical traceback ages and determine whether observational errors could account for the $\Delta_{\rm Age}$. We used as a centroid the present positions and velocities of $\beta$~Pic, and we traced this centroid back in time using \texttt{Galpy} and the \texttt{MWPotential14} Galactic potential \cite{Bovy15}. We generated 500 particles around the centroid with an initial dispersion in positions $\sigma_\textup{pos}=10$~pc and an initial dispersion in velocities $\sigma_\textup{vel}=2$~km/s. We integrated all the particles forwards in time to the present configuration, where we convolved the \textit{true} positions and velocities with realistic observational uncertainties from \textit{Gaia} plus precise spectroscopic radial velocities, $\sigma_\textup{err, pos}=0.1$~pc, $\sigma_\textup{err, vel}=0.5$~km/s. Finally, we integrated the orbits of all the particles back in time twice the age of the association and determined the dynamical traceback age \cite{Miret-Roig+2018, Miret-Roig+2020b}. We repeated this process 1\,000 times and computed the mean and the standard deviation of the dynamical traceback age. For associations younger than 40~Myr, the difference between the \textit{true} age in the simulation and the estimated dynamical traceback age is of $\lesssim 1$~Myr (see the blue shaded area in Fig.~\ref{fig:deltaT}). This is similar to the observational biases in dynamical traceback ages recently measured in $\beta$~Pic \cite{Couture+2023}. 

\noindent{\bf \large Measuring $\Delta_{\text{Age}}$ }

The associations considered in this study are well-known, and many previous works determined ages from evolutionary models using different methodologies, models (mainly with the PARSEC isochrones, especially the recent works, but not all the studies considered), and stellar censuses (see Supplementary Table~1). To test the robustness of our results, we took three reference samples to compute the $\Delta_{\text{Age}}$.
\begin{itemize}
    \item \textbf{All, $\Delta_{\rm Age~All}$}. This sample includes all the previous age determinations from evolutionary models with an uncertainty of less than 5~Myr. It is a heterogeneous sample regarding methodology, evolutionary models, and association membership.
    \item \textbf{Isochronal ages (this work), $\Delta_{\rm Age~Isoc}$}. This sample only includes the ages from isochrone fitting, with the PARSEC v1.2S evolutionary models determined in this work. These ages have the advantage of being homogeneous in technique and evolutionary models, and they use the same sample of stars used for the dynamical traceback ages. However, they do not account for systematic biases among different evolutionary models. With non-standard models (i.e., models that incorporate additional surface physics such as magnetic fields and/or rotation), the differences between evolutionary models could be larger \cite{Jackson+2013, Somers+2014}.
    \item \textbf{Lithium ages, $\Delta_{\rm Age~Lit}$}. Using only the lithium ages, we calculated the mean $\Delta_{\text{Age}}$. In principle, this sample is less model-dependent, although significant age differences have been found depending on the evolutionary models considered \cite{Galindo-Guil+2022}. This sample is homogeneous in methodology but heterogeneous in the evolutionary models used and association membership.
\end{itemize}

We computed the mean $\Delta_{\rm Age}$ for each association and sample and then averaged the values from different associations to obtain the final $\langle\Delta_{\rm Age}\rangle$ for each of the three reference samples (see Table~\ref{tab:properties}). These three values are compatible within $1\sigma$ uncertainties, however, some differences are noticeable. The homogenous comparison between isochronal ages and dynamical traceback ages, $\langle\Delta_{\rm Age~Isoc}\rangle$, gives the smallest value, although not consistent with a null $\Delta_{\rm Age}$, supporting the conclusions of this study. The comparison with Lithium ages, $\langle\Delta_{\rm Age~Lit}\rangle$, is the largest, although only available for the two oldest associations in our sample, $\beta$~Pic and Tuc-Hor. For further discussion in the text, we take the $\langle\Delta_{\rm Age~All}\rangle$ as the reference, as it is more representative of the biases of different age determinations.

$\rho$~Oph is the youngest association included in this study, for which we could not determine a dynamical traceback age. The $\Delta_{\text{Age}}$ determined in this group may be a lower limit if the association has not started to expand yet. Therefore, we also computed the $\langle\Delta_{\text{Age}}\rangle$ excluding $\rho$~Oph (see Table~\ref{tab:properties}). These values are decimals of Myr older than the result we obtained including this group, suggesting that $\rho$~Oph is very close to starting free expansion. This is also observable in Figure~\ref{fig:deltaT}, where the $\Delta_{\text{Age}}$ of $\rho$~Oph is within the scatter of the rest of the groups.

\noindent{\bf \large Modeling the relation between  $\Delta_{\text{Age}}$ and the number of association members }

We employed Bayesian linear regression analysis to investigate the relation between $\Delta_{\text{Age}}$ and the number of association members (see Fig.~\ref{fig:deltaT_numMembers}). This approach offers a robust framework by providing a distribution of plausible fitting parameters rather than a single best-fitting line. To quantify and test the correlation between these two quantities, we used a linear model $y = mx + b$, where $y$ represents $\Delta_{\text{Age}}$ and $x$ denotes the number of members per association. The slope ($m$) and intercept ($b$) of the regression line represent the parameters of interest. We assumed Gaussian deviations from the linear model with zero mean and a known $\Delta_{\text{Age}}$ variance ($\sigma$). For a given association number of members ($x_i$), $\Delta_{\text{Age}}$ value ($y_i$) and uncertainty ($\sigma_i$), listed in Table~\ref{tab:properties}, slope ($m$), and intercept ($b$), the density of observed age differences $p(y_i | x_i, \sigma_i, m, b)$ becomes:
\begin{equation}\label{eq:gauss_proba}
    p(y_i \mid x_i,\sigma_i,m,b) = \frac{1}{\sqrt{2 \pi \sigma_i^2}}\text{exp} \left( - \frac{(y_i - m x_i - b)^2}{2 \sigma_i^2} \right).
\end{equation}
We obtained the likelihood by factoring these conditional probabilities for the $N = 6$ data points:
\begin{equation}\
    \mathcal{L} = \prod_{i=1}^{N} p(y_i \mid x_i,\sigma_i,m,b),
\end{equation}
and the posterior probability density function (PDF) by applying Bayes’ theorem:
\begin{equation}
    p(m, b \mid \{y_i\}_{i=1}^{N}, I) \propto p(\{y_i\}_{i=1}^{N} \mid m, b, I)\,p(m, b \mid I).
\end{equation}

The PDF $p(m, b, f \mid I)$ describes our prior knowledge of the regression parameters $(m, b)$, while $I$ summarizes all the prior knowledge of the $x_i$ and $\sigma_i$. We employed weakly informative priors with normal densities for $m$ and $b$, with a standard deviation of $10$~Myr/member and 10~Myr, respectively. We set the mean density for the intercept ($b$) to $5.5$~Myr, which corresponds to the average $\Delta_{\text{Age}}$ across the data set, and a zero-mean prior density for the slope parameter ($m$), assuming no correlation between $x$ and $y$. 

We used the \texttt{python} code \texttt{PyMC} \cite{Salvatier+2016}, which implements the Markov Chain Monte Carlo (MCMC) methodology, to sample the posterior PDF. For each parameter $(m, b)$, we computed its maximum a posteriori (MAP) position, representing the best-fitting line plotted in Figure~\ref{fig:deltaT_numMembers}. 
The grey shaded area shows the $1 \sigma$ confidence interval within which approximately $68\%$ lines are expected to lie. The complete PDF distributions for the $(m, b)$ parameters are displayed in Supplementary Figure~2.

We computed the marginal posterior PDF of the slope parameter $p(m \mid \{y_i\}_{i=1}^{N}, I)$ to determine the likelihood of a negative correlation between the number of association members and $\Delta_{\text{Age}}$. We find a small probability (approximately $5\%$) of the slope being positive ($m > 0$):
\begin{equation}
    \int_{0}^{\infty} \,dm\,p(m \mid \{y_i\}_{i=1}^{N}, I) \approx 0.05.
\end{equation}

The Bayesian linear model substantiates a negative correlation between the $\Delta_{\text{Age}}$ and the number of members in the association. Future studies with larger samples are needed to re-evaluate this trend with higher statistical power. 

\end{methods}

\newpage
\begin{addendum}

 \item [Data availability] The data that support the findings of this study is available in the Tables in the manuscript.

 \item[Acknowledgments] We acknowledge the constructive feedback of three anonymous referees that helped to improve the clarity and quality of the manuscript. D. Barrado is supported by Spanish MCIN/AEI/10.13039/501100011033 grant PID2019-107061GB-C61 and No. MDM-2017-0737. S.~Ratzenb{\"o}ck acknowledges funding by the Austrian Research Promotion Agency (FFG, \url{https://www.ffg.at/}) under project number FO999892674.

 \item [Author Contributions] 
 N. Miret-Roig led the analysis and wrote the manuscript. S.~Ratzenb{\"o}ck computed the isochrone fitting ages reported in this study and did the Bayesian modelling in Figure~3. All authors participated in the scientific discussion.

 \item[Competing Interests] The authors declare that they have no competing financial interests.

 \item[Correspondence] Correspondence and requests for materials should be addressed to N.M.R (\url{nuria.miret.roig@univie.ac.at}).

\end{addendum}



\section*{Supplementary material}

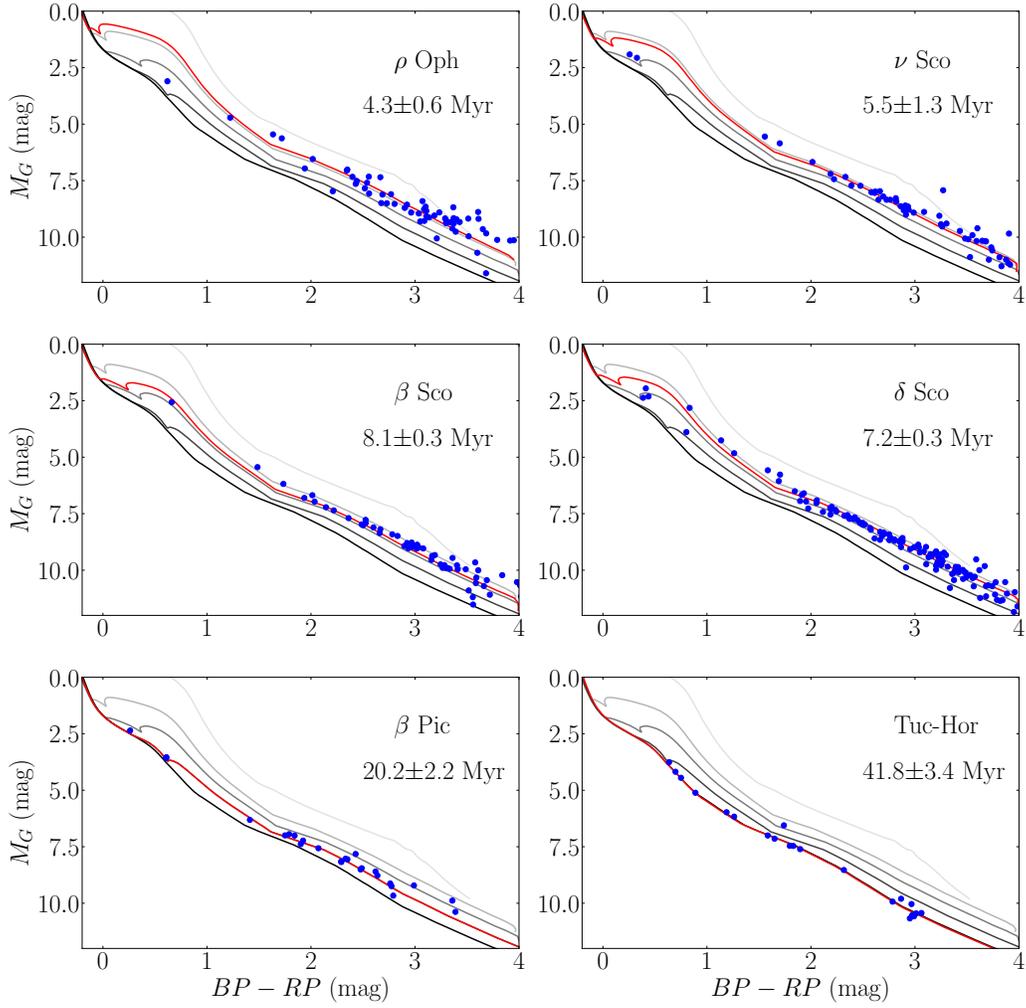

Figure 1: Colour magnitude diagram of the members of the groups considered in this study. The PARSEC isochrones corresponding to the fit computed in this work (using the algorithm[1]) are indicated in red. The PARSEC isochrones at 1, 5, 10, 20, and 40 Myr are also indicated.

1

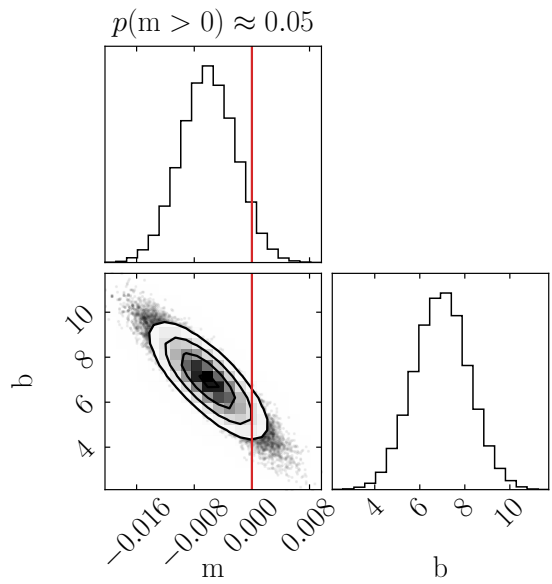

Figure 2: Corner plot showing the posterior PDF of the Bayesian model of the relation between $\Delta_{\mathrm{Age}}$ and the number of association members (Fig. 3). The fit parameters are $m$ (slope) and $b$ (intercept). Diagonal histograms display marginalised parameter distributions and the bottom left panel shows the 2D posterior distribution. The red line indicates the constant relationship ($m = 0$).

Table 1: Literature age estimates for the four regions analysed in this study.

| Region | Method | Age | Reference | Used |
|---|---|---|---|---|
| ρ Oph | Isochrone | $5.7 \pm 0.4$ | Kerr et al. (2021)[2] | Y |
| | | $3.8^{+0.4}_{-0.4}$ | Ratzenböck et al. (2023)[1] | Y |
| | | $4 \pm 1$ | Briceño-Morales et al. (2023)[3] | Y |
| | | $4.3^{+0.6}_{-0.5}$ | **This work** | Y |
| | Kinematics | $0.0 \pm 0.3$ | Miret-Roig et al. (2022)[4] | Y |
| ν Sco | Isochrone | $7.2 \pm 0.7$ | Kerr et al. (2021)[2] | Y |
| | | $5.8^{+1.8}_{-0.5}$ | Ratzenböck et al. (2023)[1] | Y |
| | | $7 \pm 1$ | Briceño-Morales et al. (2023)[3] | Y |
| | | $5.5^{+1.3}_{-0.4}$ | **This work** | Y |
| | Kinematics | $0.3 \pm 0.5$ | Miret-Roig et al. (2022)[4] | Y |
| β Sco | Isochrone | $13.2 \pm 2.8$ | Kerr et al. (2021)[2] | Y |
| | | $7.6^{+0.8}_{-0.7}$ | Ratzenböck et al. (2023)[1] | Y |
| | | $8 \pm 1$ | Briceño-Morales et al. (2023)[3] | Y |
| | | $8.1^{+0.3}_{-0.2}$ | **This work** | Y |
| | Kinematics | $2.5 \pm 1.6$ | Miret-Roig et al. (2022)[4] | Y |
| δ Sco | Isochrone | $10.2 \pm 0.7$ | Kerr et al. (2021)[2] | Y |
| | | $9.8^{+1.2}_{-1.4}$ | Ratzenböck et al. (2023)[1] | Y |
| | | $9 \pm 2$ | Briceño-Morales et al. (2023)[3] | Y |
| | | $7.2^{+0.3}_{-0.2}$ | **This work** | Y |
| | Kinematics | $4.6 \pm 1.1$ | Miret-Roig et al. (2022)[4] | Y |
| β Pic | Isochone | $20 \pm 10$ | Barrado y Navascués et al. (1999b)[5] | N |
| | | $12^{+8}_{-4}$ | Zuckerman et al. (2001)[6] | N |
| | | $21.5 \pm 6.5$ | Malo et al. (2014)[7] | N |
| | | $22 \pm 3$ | Mamajek & Bell (2014)[8] | Y |
| | | $24 \pm 3$ | Bell et al. (2015)[9] | Y |
| | | $5.5 - 54.5$ Myr | Ujjwal et al. (2020)[10] | N |
| | | $20.2^{+2.2}_{-1.9}$ | **This work** | Y |
| | Kinematics | $11.5$ | Ortega et al. (2002)[11] | N |
| | | $12$ | Song et al. (2003)[12] | N |
| | | $10.8 \pm 0.3$ | Ortega et al. (2004)[13] | N |
| | | $18$ | Torres et al. (2006)[14] | N |
| | | $31 \pm 21$ | Makarov (2007)[15] | N |
| | | $13 - 58$ | Mamajek & Bell (2014)[8] | N |
| | | $13^{+7}_{-0}$ | Miret-Roig et al. (2018)[16] | N |
| | | $17.8 \pm 1.2$ | Crundall et al. (2019)[17] | N |
| | | $18.5^{+2.0}_{-2.4}$ | Miret-Roig et al. (2020)[18] | Y |
| | | $20.4 \pm 2.5$ | Couture et al. (2023)[19] | N |
| | Lithium | $21 \pm 9$ | Mentuch et al. (2008)[20] | N |
| | | $\sim 40$ | Macdonald & Mullan (2010)[21] | N |
| | | $21 \pm 4$ | Binks & Jeffries (2014)[22] | Y |
| | | $26 \pm 3$ | Malo et al. (2014)[7] | Y |
| | | $25 \pm 3$ | Messina et al. (2016)[23] | Y |
| | | $24.3^{+0.3}_{-0.3}$ | Galindo-Guil et al. (2022)[24] | Y |
| | | $22.9^{+1.1}_{-1.0}$ | Jeffries et al. (2023)[25] | Y |
| Tuc-Hor | Isochone | $45 \pm 4$ | Bell et al. (2015)[9] | Y |
| | | $46.3 \pm 2.3$ | Kerr et al. (2022a)[26] | Y |
| | | $41.8^{+3.4}_{-2.2}$ | **This work** | Y |
| | Kinematics | $5^{+23}_{-0}$ | Miret-Roig et al. (2018)[16] | N |
| | | $32.9 \pm 8.2$ | Kerr et al. (2022a)[26] | N |
| | | $38.5^{+1.6}_{-8.0}$ | Galli et al. (2023)[27] | Y |
| | Lithium | $21 \pm 9$ | Mentuch et al. (2008)[20] | N |
| | | $\sim 40$ | Kraus et al. (2014)[28] | N |
| | | $51.0^{+0.5}_{-0.2}$ | Galindo-Guil et al. (2022)[24] | Y |
| | | $41.7^{+3.0}_{-1.9}$ | Jeffries et al. (2023)[25] | Y |

**Notes.** Only the age determinations with an uncertainty < 5 Myr were considered for this study, indicated with "Y" in the "Used" column.



\end{document}